\documentclass[%
 reprint,
 amsmath,amssymb,
 aps,
]{revtex4-2}

\usepackage{siunitx}
\usepackage{xcolor}
\usepackage{graphicx}
\usepackage{dcolumn}
\usepackage{bm}
\usepackage{hyperref}
\usepackage{physics}
\usepackage{amsmath}
\usepackage{amssymb}

\begin{document}

\preprint{APS/123-QED}
\title{Spin polarisation signatures of Fractionally Charged Skyrmions in Fractional Quantum Hall states}
\author{Odysseas Williams}
\author{Stefan Faelt}
\author{Christian Reichl}
\author{Werner Wegscheider}
\affiliation{Solid State Physics Laboratory, ETH Zurich}%
\date{\today}
\begin{abstract}
We investigate spin polarisation and low-energy excitations in fractional quantum Hall (FQH) states using cavity-polariton spectroscopy of high-mobility GaAs quantum wells. By measuring the optical coupling strength of interband Landau-level excitations over the range $1/3 \le \nu \le 1$, we extract the spin polarisation of the electron system as a function of filling factor. 

Complete suppression of the oscillator strength of the lowest energy excitation, characteristic of singlet trion formation in fully polarised systems, is reported for the first time in this regime. 

At large magnetic fields, fully polarised FQH states exhibit symmetric depolarisation away from their quantised fillings, analogous to Skyrmionic behaviour near $\nu=1$. The depolarisation follows an empirical law $S=\nu^*$, where $S$ is the number of spin flips per added magnetic flux quantum and $\nu^*$ the effective Composite Fermion filling factor. We interpret this behaviour as evidence for Minimal Fractionally Charged Skyrmions (MFCS) formed from bound spin-flip and quasiparticle excitations.
\end{abstract}
\maketitle


\section{Introduction}

Fractional quantum Hall (FQH) states are strongly correlated topological phases that occur at fractional filling factors $\nu$, the most experimentally accessible of which belong to the first Jain sequence,
\begin{equation}
    \nu_\pm = \frac{\nu^*}{2\nu^*\pm1}
    \label{eq:jain_formula}
\end{equation}
where $\nu^*$ is the effective integer filling factor. The ground-state wavefunction of the first member of this sequence, $\nu = 1/3$, was proposed by Laughlin in 1983 \cite{Laughlin_1983}. Jain later showed that attaching two magnetic flux quanta to each electron maps electrons at fractional filling $\nu$ onto Composite Fermions (CFs) occupying pseudo-Landau levels ($\Lambda$Ls) at an effective integer filling $\nu^*$ \cite{Jain_1989}. The resulting CF ground-state wavefunction closely resembles the Laughlin wavefunction and provides an intuitive framework for the hierarchy of observed fractions.

Despite this success, the nature of excitations in FQH states remains debated. In Laughlin’s theory, quasiparticles of states at $\nu = 1/m$ are anyons carrying fractional charge $\pm e/m$, where $e$ is the elementary charge, and obey fractional exchange statistics. These excitations are central to proposed topological quantum computing schemes \cite{Nayak_2008,Lachezar_2017} and probe fundamental aspects of topology and quantum statistics. A recent experiment directly measured the $\pi/3$ exchange statistics of the $\nu=1/3$ state \cite{Nakamura_2020}.

In CF theory, elementary excitations are naively described as single CFs occupying higher $\Lambda$Ls. However, Laughlin and CF quasiparticles differ in energy \cite{Jeon_2003}, implying residual CF interactions that generate more complex excitations \cite{Jain_2005}. Such interactions are believed to underlie more exotic FQH states, including the still poorly understood $\nu=5/2$ state \cite{Moore_1991,Kumar_2025}.

Here, we report spin-polarisation measurements on two samples in the range $1/3 \le \nu \le 1$. Spin polarisation has long served as a sensitive probe of quantum Hall states, providing information beyond conventional transport measurements. In particular, the $\nu=1$ state was shown to be ferromagnetic, with full spin polarisation $P=1$ even at low magnetic fields. The sharp, symmetric depolarisation away from $\nu=1$ provided the first experimental evidence for Skyrmions (extended spin textures) as the elementary excitations of this ferromagnetic state, rather than isolated spin flips \cite{Barrett_1995,Aifer_1996,Manfra_1997,Plochoka_2009,Lupatini_2020}.

In the FQH regime, characteristic values of $P$, associated with occupations of different $\Lambda$Ls, strongly support the CF description. Critical electron densities $n_e$ at which FQH states transition between spin polarisations provide information on excitation gaps. Although many such measurements have been reported \cite{Khandelwal_1998,Kukushkin_1999,Sasaki_2003,Plochoka_2009,Yoo_2020,Williams_2025,Groshaus_2007,Freytag_2001}, comparatively little is known about the continuous evolution of $P$ with $\nu$, and existing experiments report conflicting results. Analogous to the role of depolarisation near $\nu=1$ in identifying Skyrmions, measurements of $P$ away from fractional fillings may reveal the nature of FQH excitations. In this context, our experiments reveal consistent spin polarisation signatures of collective composite-fermion excitations across multiple FQH states and devices.

The spin polarisation of our systems is probed through its circularly polarised optical response, a relatively standard tool in the exploration of such physics \cite{Kukushkin_1999,Plochoka_2009,Williams_2025,Groshaus_2007}. A key observation in our study is the complete suppression of the lowest-energy optical coupling at several FQH states, consistent with singlet trion formation. This provides direct access to the minority-spin population and enables continuous spin-polarisation measurements.

\begin{figure*}[htb]  
    \includegraphics[width=\textwidth]{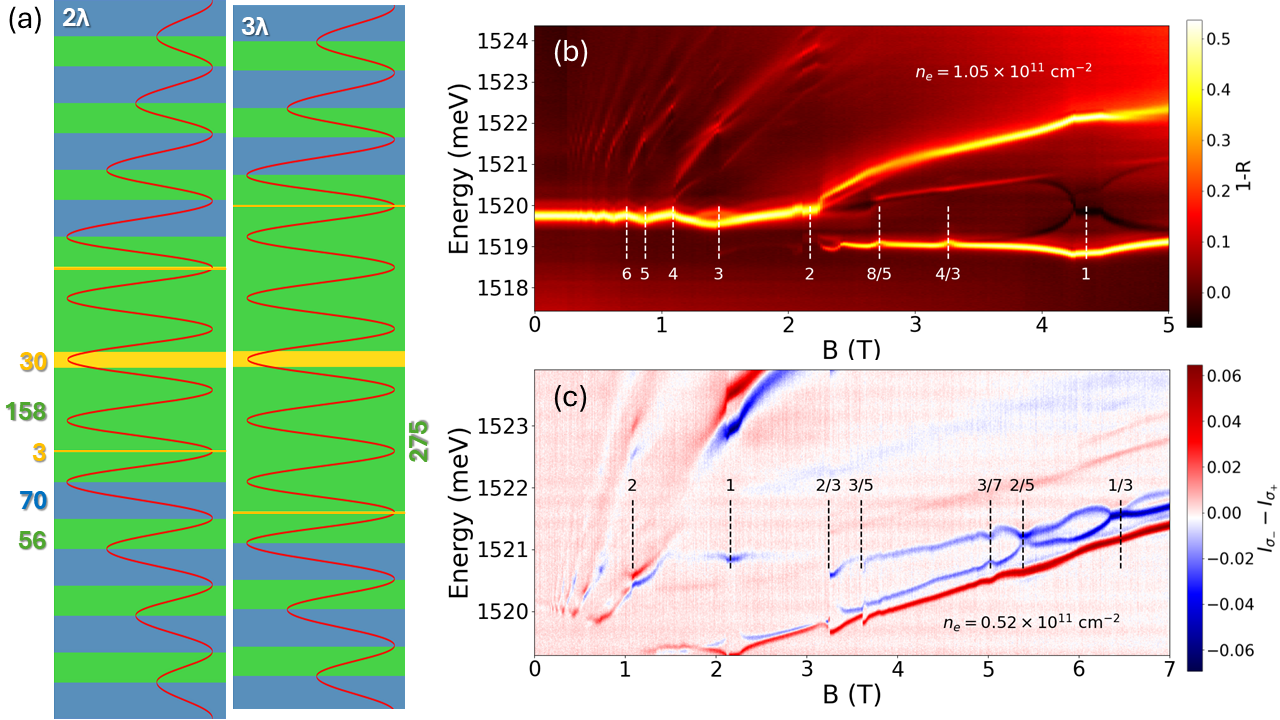}
    \caption{(a) Optical transfer-matrix simulations used to design devices $2\lambda$ and $3\lambda$. Only the layers nearest to the central cavity are shown. Blue, green, and yellow layers correspond to AlAs, Al$_{0.1}$Ga$_{0.9}$As and GaAs, respectively. The red curve corresponds to the electric field intensity of a confined resonant photon. Numbers next to each layer indicate layer thicknesses (nm); the single differing value highlights the design change in device $3\lambda$. (b) $1-R$ spectra of sample $2\lambda$ as a function of perpendicular magnetic field $B$. New polariton branches and changes in polariton splitting associated with integer and fractional $\nu$ are marked by white dashed lines. (c) Same measurement for sample $3\lambda$. Spectra measured for positive and negative $B$ were combined as $I_{\sigma_-}-I_{\sigma_+}$, allowing $\sigma_-$ polaritons (blue) and $\sigma_+$ polaritons (red) to be visualised simultaneously.}\label{fig:device_presentation}  
\end{figure*}

At low magnetic fields, we observe fractional values of $P$ for the $\nu = 2/3$, $3/5$, and $4/7$ states, together with abrupt transitions to $P=1$, consistent with partially polarised states and $\Lambda$L crossings predicted by CF theory. At higher fields, fully polarised FQH states exhibit symmetric depolarisation away from their quantised fillings, closely resembling the Skyrmionic features observed near $\nu=1$. This behaviour suggests that the relevant excitations are bound CF states. In particular, the depolarisation of the fully polarised $\nu=3/7$, $4/9$, $3/5$, and $4/7$ states is consistent with Minimal Fractionally Charged Skyrmions (MFCS), formed by binding a CF or CF hole to a spin-flip excitation. The behaviour near $\nu=1/3$, $2/3$, and $2/5$ is less clear, likely due to interplay with the larger Skyrmions of the $\nu=1$ state and higher-order fractions such as $\nu=4/11$.

\section{Methods}

The devices used in our experiments are $n$-doped Gallium Arsenide (GaAs) Quantum Wells (QW) coupled to optical microcavities. These systems host polaritons, hybrid light-matter particles, whose energies can be measured to deduce the optical coupling strength for the 2-Dimensional Electron Gas (2DEG) sitting in the QW. This coupling strength, together with selective circular polarisation of the studied light, has been widely used as a spin polarisation measurement tool, most prominently in photoluminescence and absorption studies of integer states $\nu$ \cite{Aifer_1996,Manfra_1997,Groshaus_2007,Plochoka_2009}. More recently, the usage of this technique in polaritonic devices, which allow for weaker perturbation of the system, has demonstrated its effectiveness in realising spin polarisation measurements of QH states, showing strong agreement with CF theory in the range $1 \le \nu \le 2$ \cite{Lupatini_2020,Ravets_2018,Knueppel2019,Williams_2025}.

Aluminium Arsenide (AlAs) / GaAs heterostructures are grown on a GaAs substrate using Molecular Beam Epitaxy (MBE). The exceptional quality of the MBE growth is reflected in optimised GaAs heterostructures grown in similar conditions, with electron mobilities exceeding $30\times10^6$ cm$^2$V$^{-1}$s$^{-1}$ \cite{Kulah_2021}, and is essential for resolving the delicate FQH features reported here.

Planar optical cavities with a resonant wavelength $\lambda = 815$ nm are realised by stacking two Distributed Bragg Reflectors (DBR), which consist of periodically alternating $\lambda/4n$ width layers of AlAs and Al$_{0.1}$Ga$_{0.9}$As, with $n$ the refractive index for each material, and by separating the DBRs with a $k\lambda/n$ width Al$_{0.1}$Ga$_{0.9}$As cavity, with $k$ an integer. The grown layer stacks are thicker in their center, gradually thinning out towards the edge, allowing us to tune the resonant wavelength of the device by realising optical measurements at different positions on the wafer.

A 30 nm wide GaAs QW is grown at the center of the cavity, where the electric field intensity of a confined photon is maximal, for efficient QW-photon coupling. The QW is populated with a 2DEG by placing Silicon $\delta$-doping sites at distances $k_{\text{odd}}\lambda/4$ away from the central 2DEG (with $k_\text{odd}$ an odd integer), at nodes of the electric field, where their interaction with cavity photons is strongly suppressed. We further protect them from interaction with light by placing them in 3 nm GaAs QWs, in order to reduce the formation of optically active DX centers \cite{Ababou_1990,Miwa1999DX}.

Two devices, denoted $2\lambda$ and $3\lambda$, were fabricated (Fig.~\ref{fig:device_presentation}(a)). Both contain 22.5 DBR periods on the substrate side and 19 periods on the vacuum side. Their only difference is the cavity length: sample $2\lambda$ has a $2\lambda$ cavity with doping QWs located $3\lambda/4$ from the central QW, whereas sample $3\lambda$ has a $3\lambda$ cavity with doping QWs at $5\lambda/4$. Consequently, sample $2\lambda$ has a higher electron density, $n_{e,2\lambda} = 1.05\times10^{11}$ cm$^{-2}$, compared with $n_{e,3\lambda} = 0.52\times10^{11}$ cm$^{-2}$ for sample $3\lambda$.

\begin{figure}[htb]  
    \includegraphics[width=3.4in]{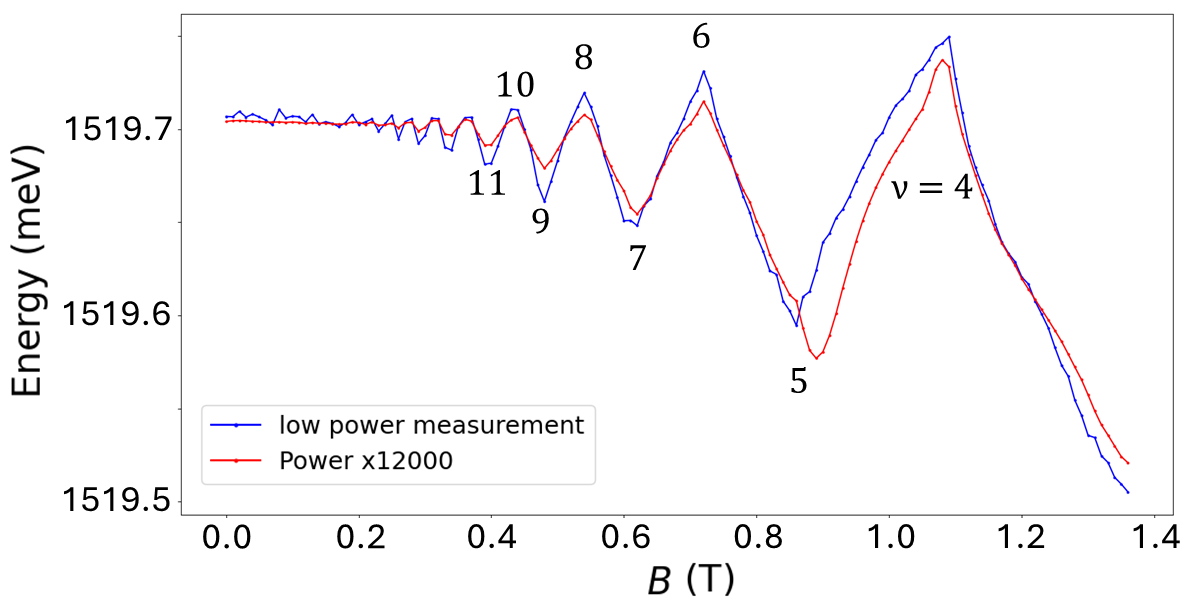}
    \caption{Lowest polariton energy in sample $2\lambda$ as a function of magnetic field $B$, measured at low (blue) and high (red) optical powers differing by a factor of $1.2\times10^4$. Integer filling factors corresponding to maxima and minima are labelled in black.}\label{fig:light_sensitivity}  
\end{figure}

The electron densities were determined from normal-incidence reflection spectra measured at fixed spatial positions while sweeping the perpendicular magnetic field $B$ at sub-Kelvin temperatures (Fig.~\ref{fig:device_presentation}(b,c)). Discrete changes in the optical response were identified with integer and fractional QH states at filling factors $\nu$, yielding the density through
\begin{equation}
    B_\nu = \frac{hn_e }{e\nu}
\end{equation}
where $h$ is Planck’s constant and $e$ the elementary charge. Measurements repeated over more than four orders of magnitude in optical power showed no measurable change in $n_e$, demonstrating negligible light sensitivity. Figure \ref{fig:light_sensitivity} compares measurements on sample $2\lambda$ performed at low and high optical powers: all integer filling factors with $\nu>5$ occur at identical magnetic fields, indicating constant density. The small discrepancy at $\nu=5$ is instead attributed to nonlinear optical effects in the QW excitations, which also reduce the oscillation amplitude at high power. Single-photon nonlinearities in FQH states were previously reported in \cite{Knueppel2019}, while continuous-power-induced polariton shifts at $\nu=1$ were observed in \cite{Williams_2025}.

Apart from the device characterisation described above, the measurement protocol follows that of Ref.~\cite{Williams_2025}. Sample $2\lambda$ was cooled to 250 mK in a $^3$He cryostat, while sample $3\lambda$ was cooled to 25 mK in a $^3$He/$^4$He dilution refrigerator.

Normal-incidence reflection spectra were measured while sweeping both the perpendicular magnetic field $B$ and the cavity energy $E_C$. Measurements were repeated for both circular polarisations by reversing the sign of $B$. The spectra were fitted with a coupled-oscillator model to extract excitation energies $E_n$ and Rabi splittings $\Omega_n$, which characterise the coupling strength between the cavity mode $E_C$ and the matter excitation $E_n$.

\section{Results and discussion}

\subsection{Excitation and coupling energies}

Excitation and coupling energies of interband Landau Level (LL) transitions were measured for 2 T$<B<$ 8T over the two samples, giving access to the filling factor ranges $1>\nu>1/2$ for sample $2\lambda$ and $1>\nu>1/3$ for sample $3\lambda$.

\begin{figure*}[htb]  
    \includegraphics[width=\textwidth]{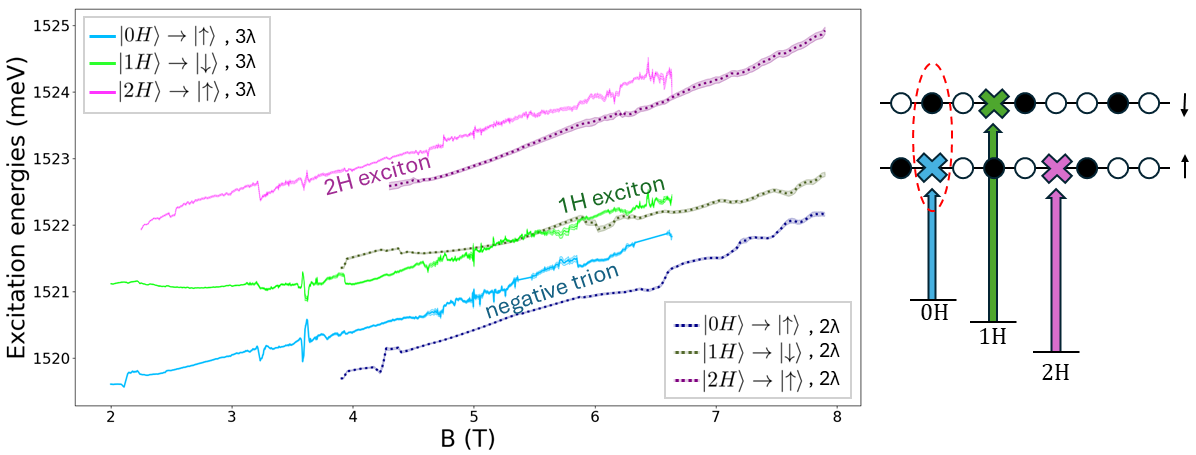}
    \caption{Three lowest excitation energies $E_0$, $E_1$, and $E_2$ measured in $\sigma_-$ polarisation for sample $2\lambda$ in the range 4 T $<B<$ 8 T and sample $3\lambda$ in the range 2 T $<B<$ 6.7 T. The schematic on the right illustrates the proposed interpretation: each excitation originates from a different hole subband ($0H$, $1H$, or $2H$) and forms an exciton (cross symbol) whose electron component occupies one of the two spin-split lowest LLs. Filled circles denote background electrons and empty circles holes; the illustrated configuration corresponds to an unpolarised $\nu=2/3$ state. The lowest-energy excitation forms a negative trion, in which a $\ket{\uparrow}$ exciton binds a $\ket{\downarrow}$ electron.}\label{fig:excitation_energies}  
\end{figure*}

\begin{figure}[htb]
    \includegraphics[width=3.4in]{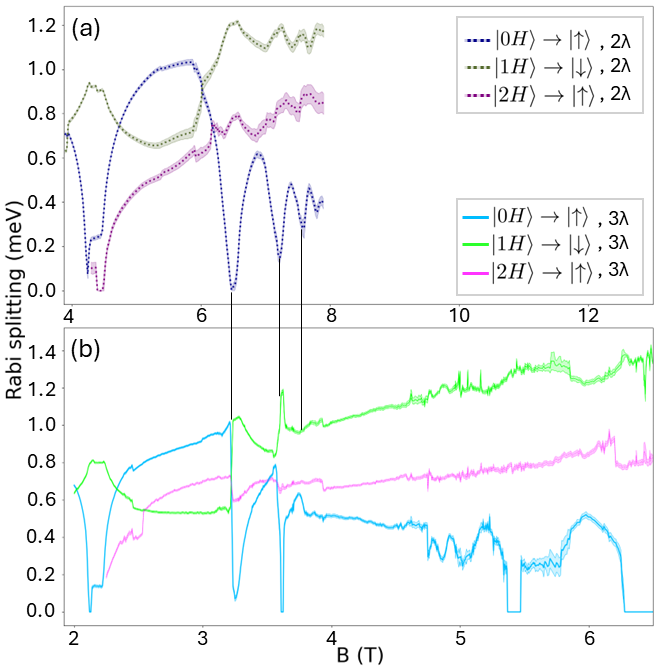}
    \caption{Coupling energies $\Omega_0$, $\Omega_1$, and $\Omega_2$ measured for samples (a) $2\lambda$ and (b) $3\lambda$. These values are extracted from the same measurements as Fig.~\ref{fig:excitation_energies}, with matching colour schemes between figures. Vertical black lines indicate filling factors $\nu=2/3$, $3/5$, and $4/7$ (from left to right).}\label{fig:Rabi_splittings}  
\end{figure}

The three lowest excitation energies measured in $\sigma_-$ polarisation are shown in Fig.~\ref{fig:excitation_energies}. The two datasets were acquired approximately one year apart without spectrometer recalibration, so their absolute energies should be interpreted cautiously. Their close agreement nevertheless indicates that they correspond to the same valence-to-conduction-band transitions.

The corresponding coupling energies are shown in Fig.~\ref{fig:Rabi_splittings}. The magnetic-field axes are scaled such that a vertical line corresponds to the same filling factor $\nu$ in both samples.

The most striking feature of these plots is that the coupling energy for the lowest energy excitation (dark and light blue) vanishes at specific values of $B$ which correspond to integer and fractional filling factors. This behaviour had already been observed for the $\nu = 1$ state, and had been attributed to the state being ferromagnetic. Since, this lowest energy transition corresponds to excitations into the conduction band $\ket{\uparrow}$ spin state, and since this state is full at $\nu=1$, no transitions can occur and the coupling vanishes.

To our knowledge, the same behaviour occuring at fractional fillings has not been reported before. While filling up the $\ket{\uparrow}$ spin state in the $\nu<1$ regime could result in a partial reduction of the optical coupling, and such observations have been made before \cite{Ravets_2018,Knueppel2019,Plochoka_2009}, its complete suppression is a new phenomenon. We attribute this improvement to the high quality of the present MBE-grown samples, as well as the high sensitivity of polariton spectroscopy. The vanishing coupling indicates that the optical excitation is a many-body bound state rather than a simple interband transition.

Interband excitations in GaAs QWs are known to form excitons, bound electron-hole pairs, with energies reduced below the bare band gap by the exciton binding energy \cite{Bastard_1982,WHITTAKER19881,Smith_1989,Potemski_1991}. In $n$-doped GaAs QWs, the situation is more complex because the photo-generated hole can bind not only to the excited electron but also to electrons already present in the conduction band.

This effect is particularly important at high magnetic fields, where different regimes emerge depending on the 2DEG density. At low densities, the excitation is commonly described as a negative trion, consisting of a hole bound to two conduction-band electrons. At higher densities, it is more naturally described as an attractive polaron, in which an exciton is dressed by excitations of the surrounding Fermi sea \cite{Shields_1995,Finkelstein_1995,Efimkin_2018,COMBESCOT2008215}. The microscopic nature of these states remains an active topic of research.

Our interpretation of the optical transitions is illustrated schematically in Fig.~\ref{fig:excitation_energies}. Since all three transitions display the same linear dependence on $B$, they must all terminate in the lowest conduction-band LL ($L=0$), whose electron cyclotron energy dominates the field dependence.

The initial states originate from different valence-band subbands arising from heavy-hole/light-hole mixing. To avoid unnecessary detail, we label the three highest-energy valence states $\ket{0H}$, $\ket{1H}$, and $\ket{2H}$. The observed transitions then correspond, from lowest to highest energy, to
\begin{equation*}
0:\ket{0H}\rightarrow\ket{\uparrow},\;\;\;\;1:\ket{1H}\rightarrow\ket{\downarrow},\;\;\;\;2:\ket{2H}\rightarrow\ket{\uparrow}.
\end{equation*}
This assignment explains why the coupling strengths of transitions 0 and 2 vanish at $\nu=1$ ($B=4.32$ T for sample $2\lambda$ and $B=2.16$ T for sample $3\lambda$; Fig.~\ref{fig:Rabi_splittings}), while transition 1 reaches a maximum, consistent with previous studies \cite{Aifer_1996,Williams_2025}.

The key new element of our interpretation is that excitation 0 forms singlet trions, which dominate the coupling behaviour in the $\nu<1$ regime. In Fig.~\ref{fig:excitation_energies}, these are represented as excitons (cross symbols) bound to opposite-spin electrons (red dashed lines). We speculate that transitions 1 and 2 are more weakly bound because the $\ket{1H}$ and $\ket{2H}$ states are less localised, resulting in smaller coupling variations. Formation of the trionic bound state in transition 0 requires the two conduction-band electrons to have opposite spin due to the Pauli exclusion principle. In strongly polarised regimes, where $\ket{\uparrow}$ electrons dominate and $\ket{\downarrow}$ electrons are scarce, this excitation is therefore better described as a singlet trion than as an attractive polaron, which would require coupling to a dense Fermi sea. Similar interpretations were previously proposed in related optical studies \cite{Groshaus_2007,Ravets_2018}. However, the complete suppression of the coupling strength at fractional fillings, which directly demonstrates the existence of these bound states, has not previously been reported.

Previous optical studies interpreted the lowest-energy coupling through $\Omega_0^2\propto n_e-n_\uparrow$, corresponding to the density of empty $\ket{\uparrow}$ states, and used this relation to extract the spin polarisation of the $\nu=1$ state \cite{Plochoka_2009,Williams_2025}. In the trionic picture relevant for $\nu<1$, however, the relevant quantity is instead the density of occupied $\ket{\downarrow}$ states, giving $\Omega_0^2\propto n_\downarrow$. The regions where $\Omega_0=0$ in Fig.~\ref{fig:Rabi_splittings} therefore correspond to fully polarised QH states with $n_\downarrow=0$.

\subsection{Spin polarisation calculation and $\Lambda$L crossing}
\label{sec:spin_pol_level_crossing}

To determine the spin polarisation across the full parameter range, the standard approach is to measure the coupling strengths in both circular polarisations and combine the results. In our case, however, this method becomes unreliable because the two lowest-energy transitions in $\sigma_+$ polarisation, clearly resolved at $\nu=1$, become nearly degenerate at larger $B$, making the field dependence of $\Omega_{\sigma_+}$ ambiguous. We therefore adopt the following assumptions:
\begin{enumerate}
    \item For $\nu<1$, the coupling strength of trion excitation 0 is given by
    \begin{equation}
    \Omega_0 = \mathcal{E}\mu\sqrt{n_\downarrow},
    \label{eq:coupling_energy}
    \end{equation}
    where $\mathcal{E}$ is the average electric field generated in the QW by a confined cavity photon, $\mu = \bra{1}-er\ket{0}$ is the dipole matrix element between the ground state $\ket{0}$ and excited state $\ket{1}$ projected onto the cavity plane, and $n_\downarrow$ is the density of occupied $\ket{\downarrow}$ states. This expression is motivated by the Jaynes–Cummings description of polaritons \cite{Imamoglu_2021}. \label{ass:jayne_cum}
    \item Variations of $\mathcal{E}$ arising from differences between the $2\lambda$ and $3\lambda$ cavity modes, as well as from changes in the photon–trion overlap with magnetic field, are negligible within our experimental accuracy. We therefore take $\mathcal{E}$ to be constant. \label{ass:E_dependence}
    \item The magnetic-field dependence of the dipole moment follows that of a single-particle LL interband transition,
    \begin{equation}
        \mu\propto l_B = \sqrt{\frac{\hbar}{eB}},
    \end{equation}
    where $l_B$ is the magnetic length, $\hbar$ the reduced Planck constant, and $e$ the elementary charge. \label{ass:mu_dependence}
    
    \item The sharp discontinuity of $\Omega_0$ at $\nu=2/3$ in sample $3\lambda$ (light blue curve in Fig.~\ref{fig:Rabi_splittings}(b), near $B=3.2$ T) corresponds to a $\Lambda$L crossing within CF theory, driving the system from an unpolarised state with $P=0$ ($n_\downarrow=n_e/2$) to a fully polarised state with $P=1$ ($n_\downarrow=0$). \label{ass:zero_polarisation}
\end{enumerate}

The spin polarisation is defined as
\begin{equation}
P = \frac{n_\uparrow - n_\downarrow}{n_\uparrow+n_\downarrow}
= 1-2\frac{n_\downarrow}{n_e},
\end{equation}
where $n_\uparrow$ and $n_\downarrow$ are the densities of electrons occupying the $\ket{\uparrow}$ and $\ket{\downarrow}$ states, respectively, and $n_e=n_\uparrow+n_\downarrow$ is the total electron density.

Using assumption \ref{ass:jayne_cum}, the spin polarisation becomes
\begin{equation}
P = 1-2\frac{\Omega_0^2}{\mathcal{E}^2\mu^2n_e}.
\end{equation}

At $B=3.2$ T in sample $3\lambda$, we define $C_0B=\mathcal{E}^2\mu^2n_{e,3\lambda}$ and determine its numerical value by imposing $P=0$ according to assumption \ref{ass:zero_polarisation}. Assumptions \ref{ass:E_dependence} and \ref{ass:mu_dependence} then yield the continuous relation
\begin{equation}
P(B)=1-2\frac{\Omega_0^2}{C_0B},
\label{eq:spin_pol_formula}
\end{equation}
which is used to extract $P(B)$ for sample $3\lambda$. For sample $2\lambda$, an additional correction factor $n_{e,2\lambda}/n_{e,3\lambda}=2$ is applied.

\begin{figure}[htb]
    \includegraphics[width=3.4in]{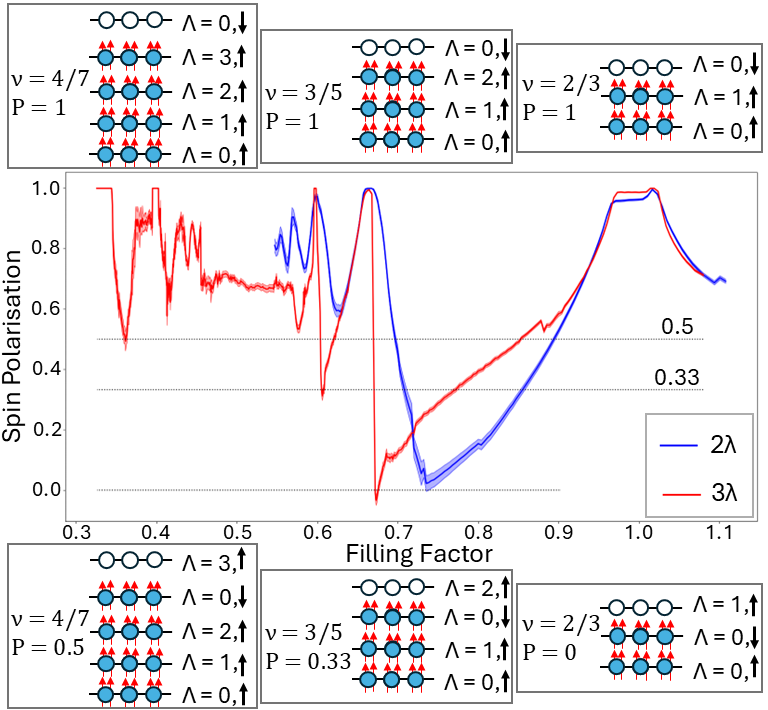}
    \caption{Spin polarisation $P$ for samples $2\lambda$ (blue) and $3\lambda$ (red), extracted from the $\Omega_0$ measurements of Fig.~\ref{fig:Rabi_splittings} using Eq.~\ref{eq:spin_pol_formula}. The accompanying diagrams show the corresponding CF configurations: top panels represent fully polarised states, bottom panels partially polarised states. The expected partial polarisations of the $\nu=3/5$ and $4/7$ states, $P=1/3$ and $1/2$, are indicated by black dotted lines.}\label{fig:spin_polarisation_full}  
\end{figure}

The resulting spin polarisation as a function of filling factor is shown in Fig.~\ref{fig:spin_polarisation_full}. For sample $3\lambda$, the discontinuous transition from $P=0$ to $P=1$ at $\nu=2/3$ is imposed by construction through assumption \ref{ass:zero_polarisation}.

A second transition is observed at $\nu=3/5$, where the polarisation jumps from $P=1/3$ to $P=1$. This behaviour is consistent with a CF $\Lambda$L crossing: for $\nu>3/5$, the $\ket{\Lambda=0,\uparrow}$, $\ket{\Lambda=1,\uparrow}$, and $\ket{\Lambda=0,\downarrow}$ levels are occupied, whereas for $\nu<3/5$, the $\ket{\Lambda=2,\uparrow}$ level drops below $\ket{\Lambda=0,\downarrow}$, causing all CFs to transfer into the spin-up branch. A $P=1/3$ plateau at $\nu=3/5$ has also been reported experimentally \cite{Kukushkin_1999,Sasaki_2003}.

Similarly, the partially polarised $\nu=4/7$ state in sample $3\lambda$, with $P=1/2$, agrees with both CF theory [bottom-left schematic in Fig.~\ref{fig:spin_polarisation_full}] and previous measurements. The quantitative agreement between these extracted polarisations, CF theory, and earlier experiments strongly supports the validity of our method for determining $P$.

\begin{figure}[htb]
    \includegraphics[width=3.4in]{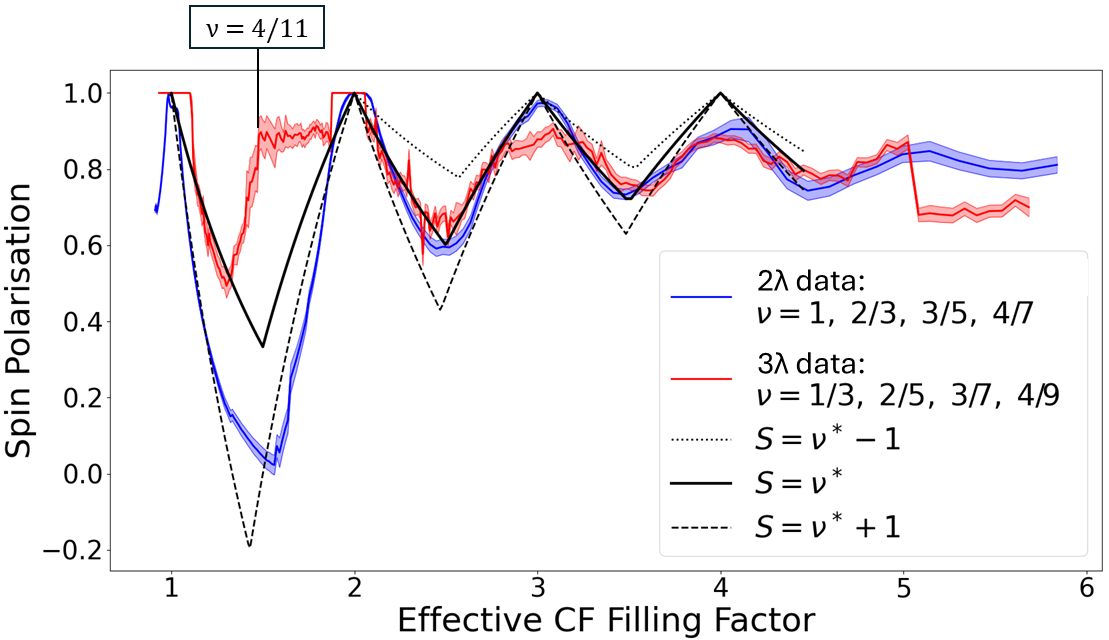}
    \caption{Spin polarisation $P$ as a function of the effective filling factor $\nu^*$ for the CF integer states. The three black lines are calculated by assuming a constant number of spin flips $S$ per added magnetic flux quantum away from a fully polarised state at integer $\nu^*$ (Eq. \ref{eq:skyrmion_polarisation}).}\label{fig:CF_spin_polarisation}  
\end{figure}

\subsection{Fully polarised states and Skyrmion formation}

Figure \ref{fig:CF_spin_polarisation} shows the same spin-polarisation data replotted as a function of the effective CF filling factor $\nu^*$, obtained from Eq.~\ref{eq:jain_formula}. The $\nu_+=1/3$, $2/5$, $3/7$, $4/9$, and $5/11$ states ($B=6.48$, 5.40, 5.04, 4.86, and 4.75 T) are shown for sample $3\lambda$ (red), while the $\nu_-=1$, $2/3$, $3/5$, $4/7$, and $5/9$ states ($B=4.32$, 6.48, 7.20, 7.56, and 7.78 T) are shown for sample $2\lambda$ (blue).

The full polarisation of these FQH states is consistent with the $\Lambda$L-crossing picture discussed above for the $\nu=2/3$, $3/5$, and $4/7$ states in sample $3\lambda$.

The energy separation between the CF states $\ket{\Lambda=0,\uparrow}$ and $\ket{\Lambda=1,\uparrow}$ is determined by the CF effective mass $m^*$. For instance, the $\nu_+=2/5$ and $\nu_-=2/3$ states in sample $3\lambda$ both correspond to an effective filling $\nu^*=2$. Since the electron density $n_e$ is fixed, the corresponding CF mass and cyclotron gap are expected to be identical \cite{Villegas_Rosales_2022,Kukushkin_2003,Nicholas_1996,Chughtai_2001}. 

In contrast, the spin splitting is governed primarily by the Zeeman energy $E_Z\propto B$, giving
\begin{equation}
\frac{E_{Z,2/5}}{E_{Z,2/3}}
=
\frac{B_{2/5}}{B_{2/3}}
=
\frac{5}{3}.
\end{equation}
Since a $\Lambda$L crossing between opposite-spin states occurs at $\nu=2/3$, the larger Zeeman energy at $\nu=2/5$ places the system well beyond the crossing regime, implying full spin polarisation. The same reasoning applies to the higher-order states and naturally explains the observed fully polarised ground states as a consequence of Zeeman-energy dominance.

The smooth depolarisation away from these fully polarised states, however, is incompatible with a purely Zeeman-dominated picture. Following the phenomenology used to describe Skyrmions near $\nu=1$ \cite{Barrett_1995,Aifer_1996,Manfra_1997,Plochoka_2009,Lupatini_2020}, we parameterise the depolarisation in terms of $S$, the number of flipped spins per added magnetic flux quantum $\phi=h/e$. Starting from a fully polarised state at filling $\nu_0$, the polarisation is written as
\begin{equation}
P_{\pm}(\nu)
=
1\pm2S\left(\frac{1}{\nu_0}-\frac{1}{\nu}\right),
\label{eq:skyrmion_polarisation}
\end{equation}
assuming $S$ remains constant, with the sign of $P_\pm$ corresponding to the sign of $\nu-\nu_0$.

In Fig.~\ref{fig:CF_spin_polarisation}, three model curves corresponding to $S=\nu^*-1$, $S=\nu^*$, and $S=\nu^*+1$ are plotted for each integer $\nu^*$. Remarkably, the depolarisation measured in both samples closely follows the $S=\nu^*$ curve over the range $2\le \nu^* \le4$.

\begin{figure}[htb]
    \includegraphics[width=3.4in]{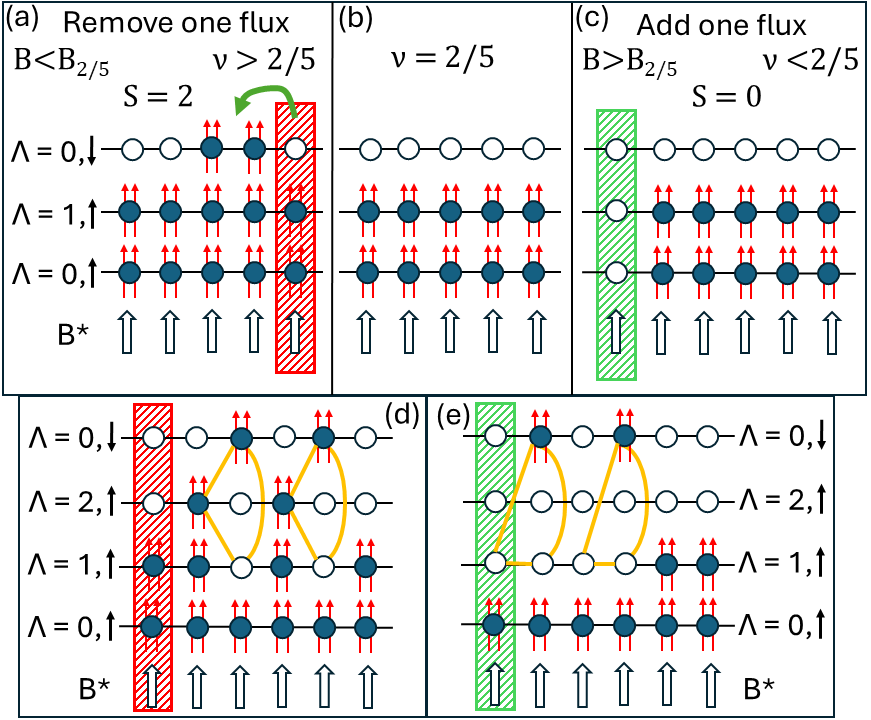}
    \caption{CF representation of the fully polarised $\nu=2/5$ ($\nu^*=2$) ground state (b). Each $\Lambda$L contains a number of states equal to the number of effective magnetic flux quanta (white arrows). The two lowest-energy levels, $\ket{\Lambda=0,\uparrow}$ and $\ket{\Lambda=1,\uparrow}$, are fully occupied by CFs (blue disks with red arrows), while higher levels remain empty (white disks). Panels (c) and (e) illustrate possible excitations obtained by adding one flux quantum (green frame), with panel (e) including an additional CF binding energy (yellow lines). Panels (a) and (d) show the corresponding excitations generated by removing one flux quantum, forcing CFs occupying the removed states to relocate.
}\label{fig:CF_excitations}  
\end{figure}

In some cases, an $S=\nu^*$ dependence could arise from elementary excitations involving a single CF occupying a higher-energy spin-reversed $\Lambda$L. Figure \ref{fig:CF_excitations}(a) illustrates this scenario for $\nu>2/5$. Removing one flux quantum from the $\nu^*=2$ ground state reduces the degeneracy of each $\Lambda$L by one state, forcing one CF per occupied $\Lambda$L into the next available level. If the lowest available state is $\ket{\Lambda=0,\downarrow}$, then two spin flips occur, giving $S=2$. The same argument generalises to fillings slightly above any integer $\nu^*$.

However, this interpretation is inconsistent with the overall behaviour of the data. If both $\nu=2/5$ and $\nu=3/7$ are fully polarised, while $\ket{\Lambda=0,\downarrow}$ remains the lowest unoccupied level at $\nu=2/5$, then a $\Lambda$L crossing must occur between these states, which should produce a sharp polarisation transition similar to those observed at lower magnetic fields (Section \ref{sec:spin_pol_level_crossing}).

Moreover, because the measurements discussed here are performed at larger magnetic fields than those where $\Lambda$L crossings were directly observed, the $\ket{\Lambda=0,\downarrow}$ level is expected to lie well above the crossing regime. Single-CF excitations should therefore not populate this state.

A complementary argument follows from considering fillings slightly below an integer $\nu^*$, illustrated in Fig.~\ref{fig:CF_excitations}(c). Here, one flux quantum is added rather than removed. If a single-CF excitation into a spin-reversed level were the dominant mechanism, then the corresponding hole excitation should involve removing a single CF from an occupied level, which would give $S=0$ and leave the system fully polarised. This is incompatible with the observed symmetric depolarisation.

We therefore interpret the observed $S=\nu^*$ behaviour on both sides of fully polarised integer $\nu^*$ states as evidence for Minimal Fractionally Charged CF Skyrmions (MFCS). In this picture, the excitation consists of a spin-flip exciton — a hole in the highest occupied $\Lambda$L bound to a CF in $\ket{\Lambda=0,\downarrow}$ — which itself binds either a CF particle in the lowest unoccupied $\Lambda$L (Fig.~\ref{fig:CF_excitations}(d)) or a CF hole in the highest occupied $\Lambda$L (Fig.~\ref{fig:CF_excitations}(e)).

An exact microscopic description of these multi-CF bound states is beyond the scope of this work. Spin-reversal excitations in fully polarised FQH states are known to arise from a complex interplay among several spin-reversed $\Lambda$Ls \cite{Wurstbauer_2011}. Measurements of the $\nu=1/3$ excitation spectrum are consistent with MFCS formation, and detailed assignments of the occupied $\Lambda$Ls have been proposed \cite{Balram_2015}. Here, we can only conclude that each excitation involves one spin flip and interpret the resulting depolarisation in terms of MFCS formation, without attempting a detailed identification of the participating $\Lambda$Ls.

This interpretation is particularly compelling for the $\nu^*=3$ and $\nu^*=4$ states, around which the depolarisation is nearly symmetric. It also naturally explains the smooth evolution between neighbouring fully polarised states. At half-integer $\nu^*$, spin-flip excitons can bind equally to CF particles and CF holes because both are present in equal numbers. Below half filling they preferentially bind CF particles, whereas above half filling they bind CF holes.

As shown in Fig.~\ref{fig:CF_spin_polarisation}, however, this picture breaks down in the range $1<\nu^*<2$. For sample $3\lambda$ (red curve), the polarisation remains large below $\nu=2/5$ before dropping sharply at $\nu=4/11$, suggesting that this higher-order state disrupts the MFCS picture.

For sample $2\lambda$ (blue curve), the depolarisation between $\nu=1$ and $\nu=2/3$ more closely follows the $S=\nu^*+1$ curve. We attribute this to the influence of the well-established integer-charge Skyrmions of the $\nu=1$ state, which involve larger spin textures ($S>1$) and therefore affect the crossover into the $\nu=2/3$ regime.

\section{Conclusion}

In conclusion, we have measured the optical coupling strength $\Omega$ in two FQH systems. The complete suppression of the coupling to the lowest-energy excitation for $\nu<1$ can only be understood in terms of bound negative trions, providing direct evidence for this type of optical excitation in this regime.

Using a set of theoretical approximations, we quantitatively extract the spin polarisation $P$ of the electron system. For sample $3\lambda$, the measured polarisations of the $\nu=3/5$ and $\nu=4/7$ states agree with previous experiments, supporting the validity of our approach. For sample $2\lambda$, the $\nu=2/3$, $3/5$, and $4/7$ states, and for sample $3\lambda$, the $\nu=2/5$, $3/7$, and $4/9$ states, are found to be fully polarised, consistent with the CF description of these FQH states.

Away from these fully polarised states, we observe a symmetric depolarisation on both sides of the corresponding integer CF fillings. Empirically, the depolarisation follows the relation $S=\nu^*$, where $S$ is the number of spin flips per added magnetic flux quantum. Importantly, this observation is reproduced across multiple FQH states and devices.

The observed scaling is inconsistent with a description based solely on independent CF quasiparticles and instead points to the importance of collective spin-reversal excitations. These results establish cavity-polariton spectroscopy as a powerful probe of FQH spin physics and provide new insight into the nature of CF excitations beyond the single-quasiparticle picture.

\section{Acknowledgements}

This project was funded by the Swiss National Science Foundation (SNSF) and by the Swiss National Center of Competence in Research Quantum Science and Technology, QSIT.

\bibliography{bibliography}

@article{Laughlin_1983,
  title = {Anomalous Quantum Hall Effect: An Incompressible Quantum Fluid with Fractionally Charged Excitations},
  author = {Laughlin, R. B.},
  journal = {Phys. Rev. Lett.},
  volume = {50},
  issue = {18},
  pages = {1395--1398},
  numpages = {0},
  year = {1983},
  month = {May},
  publisher = {American Physical Society},
  doi = {10.1103/PhysRevLett.50.1395},
  url = {https://link.aps.org/doi/10.1103/PhysRevLett.50.1395}
}

@article{Jain_1989,
  title = {Composite-fermion approach for the fractional quantum Hall effect},
  author = {Jain, J. K.},
  journal = {Phys. Rev. Lett.},
  volume = {63},
  issue = {2},
  pages = {199--202},
  numpages = {0},
  year = {1989},
  month = {Jul},
  publisher = {American Physical Society},
  doi = {10.1103/PhysRevLett.63.199},
  url = {https://link.aps.org/doi/10.1103/PhysRevLett.63.199}
}

@article{Moore_1991,
title = {Nonabelions in the fractional quantum hall effect},
journal = {Nuclear Physics B},
volume = {360},
number = {2},
pages = {362-396},
year = {1991},
issn = {0550-3213},
doi = {https://doi.org/10.1016/0550-3213(91)90407-O},
url = {https://www.sciencedirect.com/science/article/pii/055032139190407O},
author = {Gregory Moore and Nicholas Read},
}

@Article{Kumar_2025,
author={Kumar, Ravi
and Haug, Andr{\'e}
and Kim, Jehyun
and Yutushui, Misha
and Khudiakov, Konstantin
and Bhardwaj, Vishal
and Ilin, Alexey
and Watanabe, K.
and Taniguchi, T.
and Mross, David F.
and Ronen, Yuval},
title={Quarter- and half-filled quantum Hall states and their topological orders revealed by daughter states in bilayer graphene},
journal={Nature Communications},
year={2025},
month={Aug},
day={06},
volume={16},
number={1},
pages={7255},
issn={2041-1723},
doi={10.1038/s41467-025-62650-9},
url={https://doi.org/10.1038/s41467-025-62650-9}
}

@InProceedings{Lachezar_2017,
author="Georgiev, Lachezar S.",
editor="Tadjer, Alia
and Pavlov, Rossen
and Maruani, Jean
and Br{\"a}ndas, Erkki J.
and Delgado-Barrio, Gerardo",
title="Topological Quantum Computation with Non-Abelian Anyons in Fractional Quantum Hall States",
booktitle="Quantum Systems in Physics, Chemistry, and Biology",
year="2017",
publisher="Springer International Publishing",
address="Cham",
pages="75--94",
isbn="978-3-319-50255-7"
}

@article{Nayak_2008,
  title = {Non-Abelian anyons and topological quantum computation},
  author = {Nayak, Chetan and Simon, Steven H. and Stern, Ady and Freedman, Michael and Das Sarma, Sankar},
  journal = {Rev. Mod. Phys.},
  volume = {80},
  issue = {3},
  pages = {1083--1159},
  numpages = {0},
  year = {2008},
  month = {Sep},
  publisher = {American Physical Society},
  doi = {10.1103/RevModPhys.80.1083},
  url = {https://link.aps.org/doi/10.1103/RevModPhys.80.1083}
}

@Article{Nakamura_2020,
author={Nakamura, J.
and Liang, S.
and Gardner, G. C.
and Manfra, M. J.},
title={Direct observation of anyonic braiding statistics},
journal={Nature Physics},
year={2020},
month={Sep},
day={01},
volume={16},
number={9},
pages={931-936},
issn={1745-2481},
doi={10.1038/s41567-020-1019-1},
url={https://doi.org/10.1038/s41567-020-1019-1}
}

@article{Jeon_2003,
  title = {Nature of quasiparticle excitations in the fractional quantum Hall effect},
  author = {Jeon, Gun Sang and Jain, Jainendra K.},
  journal = {Phys. Rev. B},
  volume = {68},
  issue = {16},
  pages = {165346},
  numpages = {5},
  year = {2003},
  month = {Oct},
  publisher = {American Physical Society},
  doi = {10.1103/PhysRevB.68.165346},
  url = {https://link.aps.org/doi/10.1103/PhysRevB.68.165346}
}

@article{Jain_2005,
title = {Composite fermion theory of excitations in the fractional quantum Hall effect},
journal = {Solid State Communications},
volume = {135},
number = {9},
pages = {602-609},
year = {2005},
note = {Fundamental Optical and Quantum Effects in Condensed Matter},
issn = {0038-1098},
doi = {https://doi.org/10.1016/j.ssc.2005.04.033},
url = {https://www.sciencedirect.com/science/article/pii/S0038109805003947},
author = {J.K. Jain and K. Park and M.R. Peterson and V.W. Scarola},
}

@article{Wurstbauer_2011,
  title = {Observation of Nonconventional Spin Waves in Composite-Fermion Ferromagnets},
  author = {Wurstbauer, U. and Majumder, D. and Mandal, S. S. and Dujovne, I. and Rhone, T. D. and Dennis, B. S. and Rigosi, A. F. and Jain, J. K. and Pinczuk, A. and West, K. W. and Pfeiffer, L. N.},
  journal = {Phys. Rev. Lett.},
  volume = {107},
  issue = {6},
  pages = {066804},
  numpages = {4},
  year = {2011},
  month = {Aug},
  publisher = {American Physical Society},
  doi = {10.1103/PhysRevLett.107.066804},
  url = {https://link.aps.org/doi/10.1103/PhysRevLett.107.066804}
}

@Article{Balram_2015,
author={Balram, Ajit C.
and Wurstbauer, U.
and W{\'o}js, A.
and Pinczuk, A.
and Jain, J. K.},
title={Fractionally charged skyrmions in fractional quantum Hall effect},
journal={Nature Communications},
year={2015},
month={Nov},
day={26},
volume={6},
number={1},
pages={8981},
issn={2041-1723},
doi={10.1038/ncomms9981},
url={https://doi.org/10.1038/ncomms9981}
}

@article{Barrett_1995,
  title = {Optically Pumped NMR Evidence for Finite-Size Skyrmions in GaAs Quantum Wells near Landau Level Filling $\mathit{\ensuremath{\nu}}\phantom{\rule{0ex}{0ex}}=\phantom{\rule{0ex}{0ex}}1$},
  author = {Barrett, S. E. and Dabbagh, G. and Pfeiffer, L. N. and West, K. W. and Tycko, R.},
  journal = {Phys. Rev. Lett.},
  volume = {74},
  issue = {25},
  pages = {5112--5115},
  numpages = {0},
  year = {1995},
  month = {Jun},
  publisher = {American Physical Society},
  doi = {10.1103/PhysRevLett.74.5112},
  url = {https://link.aps.org/doi/10.1103/PhysRevLett.74.5112}
}

@article{Aifer_1996,
  title = {Evidence of Skyrmion Excitations about $\mathit{\ensuremath{\nu}}\phantom{\rule{0ex}{0ex}}=\phantom{\rule{0ex}{0ex}}1$ in $\mathit{n}$-Modulation-Doped Single Quantum Wells by Interband Optical Transmission},
  author = {Aifer, E. H. and Goldberg, B. B. and Broido, D. A.},
  journal = {Phys. Rev. Lett.},
  volume = {76},
  issue = {4},
  pages = {680--683},
  numpages = {0},
  year = {1996},
  month = {Jan},
  publisher = {American Physical Society},
  doi = {10.1103/PhysRevLett.76.680},
  url = {https://link.aps.org/doi/10.1103/PhysRevLett.76.680}
}

@article{Manfra_1997,
title = {Optical determination of the spin polarization of a quantum Hall ferromagnet},
journal = {Physica E: Low-dimensional Systems and Nanostructures},
volume = {1},
number = {1},
pages = {28-35},
year = {1997},
issn = {1386-9477},
doi = {https://doi.org/10.1016/S1386-9477(97)00006-4},
url = {https://www.sciencedirect.com/science/article/pii/S1386947797000064},
author = {M.J. Manfra and B.B. Goldberg and L. Pfeiffer and K. West},
}

@article{Plochoka_2009,
  title = {Optical Absorption to Probe the Quantum Hall Ferromagnet at Filling Factor $\ensuremath{\nu}=1$},
  author = {Plochocka, P. and Schneider, J. M. and Maude, D. K. and Potemski, M. and Rappaport, M. and Umansky, V. and Bar-Joseph, I. and Groshaus, J. G. and Gallais, Y. and Pinczuk, A.},
  journal = {Phys. Rev. Lett.},
  volume = {102},
  issue = {12},
  pages = {126806},
  numpages = {4},
  year = {2009},
  month = {Mar},
  publisher = {American Physical Society},
  doi = {10.1103/PhysRevLett.102.126806},
  url = {https://link.aps.org/doi/10.1103/PhysRevLett.102.126806}
}

@article{Lupatini_2020,
  title = {Spin Reversal of a Quantum Hall Ferromagnet at a Landau Level Crossing},
  author = {Lupatini, M. and Kn\"{u}ppel, P. and Faelt, S. and Winkler, R. and Shayegan, M. and Imamoglu, A. and Wegscheider, W.},
  journal = {Phys. Rev. Lett.},
  volume = {125},
  issue = {6},
  pages = {067404},
  numpages = {6},
  year = {2020},
  month = {Aug},
  publisher = {American Physical Society},
  doi = {10.1103/PhysRevLett.125.067404},
  url = {https://link.aps.org/doi/10.1103/PhysRevLett.125.067404}
}

@article{Khandelwal_1998,
  title = {Optically Pumped Nuclear Magnetic Resonance Measurements of the Electron Spin Polarization in GaAs Quantum Wells near Landau Level Filling Factor $\mathit{\ensuremath{\nu}}\phantom{\rule{0ex}{0ex}}=\phantom{\rule{0ex}{0ex}}\frac{1}{3}$},
  author = {Khandelwal, P. and Kuzma, N. N. and Barrett, S. E. and Pfeiffer, L. N. and West, K. W.},
  journal = {Phys. Rev. Lett.},
  volume = {81},
  issue = {3},
  pages = {673--676},
  numpages = {0},
  year = {1998},
  month = {Jul},
  publisher = {American Physical Society},
  doi = {10.1103/PhysRevLett.81.673},
  url = {https://link.aps.org/doi/10.1103/PhysRevLett.81.673}
}

@article{Kukushkin_1999,
  title = {Spin Polarization of Composite Fermions: Measurements of the Fermi Energy},
  author = {Kukushkin, I. V. and v. Klitzing, K. and Eberl, K.},
  journal = {Phys. Rev. Lett.},
  volume = {82},
  issue = {18},
  pages = {3665--3668},
  numpages = {0},
  year = {1999},
  month = {May},
  publisher = {American Physical Society},
  doi = {10.1103/PhysRevLett.82.3665},
  url = {https://link.aps.org/doi/10.1103/PhysRevLett.82.3665}
}

@article{Sasaki_2003,
title = {Spin polarization in fractional quantum Hall effect},
journal = {Surf. Sci.},
volume = {532-535},
pages = {567-575},
year = {2003},
note = {Proceedings of the 7th International Conference on Nanometer-Scale Science and Technology and the 21st European Conference on Surface Science},
issn = {0039-6028},
doi = {https://doi.org/10.1016/S0039-6028(03)00091-8},
url = {https://www.sciencedirect.com/science/article/pii/S0039602803000918},
author = {Shosuke Sasaki},
}

@Article{Yoo_2020,
author={Yoo, H. M.
and Baldwin, K. W.
and West, K.
and Pfeiffer, L.
and Ashoori, R. C.},
title={Spin phase diagram of the interacting quantum Hall liquid},
journal={Nat. Phys.},
year={2020},
month={Oct},
day={01},
volume={16},
number={10},
pages={1022-1027},
issn={1745-2481},
doi={10.1038/s41567-020-0946-1},
url={https://doi.org/10.1038/s41567-020-0946-1}
}

@Article{Williams_2025,
author={Williams, O.
and Faelt, S.
and Krizek, F.
and Wegscheider, W.},
title={Spin polarization of quantum Hall states for filling factors $1\ensuremath{\le\nu\le}2$ measured with microcavity polaritons},
journal={New J. Phys.},
year={2025},
month={Apr},
volume={27},
pages={043026},
doi={10.1088/1367-2630/adccf0},
url={https://doi.org/10.1088/1367-2630/adccf0}
}

@article{Groshaus_2007,
  title = {Absorption in the Fractional Quantum Hall Regime: Trion Dichroism and Spin Polarization},
  author = {Groshaus, J. G. and Plochocka-Polack, P. and Rappaport, M. and Umansky, V. and Bar-Joseph, I. and Dennis, B. S. and Pfeiffer, L. N. and West, K. W. and Gallais, Y. and Pinczuk, A.},
  journal = {Phys. Rev. Lett.},
  volume = {98},
  issue = {15},
  pages = {156803},
  numpages = {4},
  year = {2007},
  month = {Apr},
  publisher = {American Physical Society},
  doi = {10.1103/PhysRevLett.98.156803},
  url = {https://link.aps.org/doi/10.1103/PhysRevLett.98.156803}
}

@article{Freytag_2001,
  title = {New Phase Transition between Partially and Fully Polarized Quantum Hall States with Charge and Spin Gaps at $\mathit{\ensuremath{\nu}}\phantom{\rule{0ex}{0ex}}=\phantom{\rule{0ex}{0ex}}\frac{2}{3}$},
  author = {Freytag, N. and Tokunaga, Y. and Horvati\ifmmode \acute{c}\else \'{c}\fi{}, M. and Berthier, C. and Shayegan, M. and L\'evy, L. P.},
  journal = {Phys. Rev. Lett.},
  volume = {87},
  issue = {13},
  pages = {136801},
  numpages = {4},
  year = {2001},
  month = {Sep},
  publisher = {American Physical Society},
  doi = {10.1103/PhysRevLett.87.136801},
  url = {https://link.aps.org/doi/10.1103/PhysRevLett.87.136801}
}

@article{Ravets_2018,
  title = {Polaron Polaritons in the Integer and Fractional Quantum Hall Regimes},
  author = {Ravets, Sylvain and Kn\"uppel, Patrick and Faelt, Stefan and Cotlet, Ovidiu and Kroner, Martin and Wegscheider, Werner and Imamoglu, Atac},
  journal = {Phys. Rev. Lett.},
  volume = {120},
  issue = {5},
  pages = {057401},
  numpages = {6},
  year = {2018},
  month = {Jan},
  publisher = {American Physical Society},
  doi = {10.1103/PhysRevLett.120.057401},
  url = {https://link.aps.org/doi/10.1103/PhysRevLett.120.057401}
}

@Article{Knueppel2019,
author={Kn{\"u}ppel, Patrick
and Ravets, Sylvain
and Kroner, Martin
and F{\"a}lt, Stefan
and Wegscheider, Werner
and Imamoglu, Atac},
title={Nonlinear optics in the fractional quantum Hall regime},
journal={Nature},
year={2019},
month={Aug},
day={01},
volume={572},
number={7767},
pages={91-94},
issn={1476-4687},
doi={10.1038/s41586-019-1356-3},
url={https://doi.org/10.1038/s41586-019-1356-3}
}

@article{Ababou_1990,
    author = {Ababou, S. and Marchand, J. J. and Mayet, L. and Guillot, G. and Mollot, F.},
    title = {Characterization of DX centers in selectively doped GaAs‐AlAs superlattices},
    journal = {Applied Physics Letters},
    volume = {57},
    number = {13},
    pages = {1321-1323},
    year = {1990},
    month = {09},
    issn = {0003-6951},
    doi = {10.1063/1.103471},
    url = {https://doi.org/10.1063/1.103471},
}

@article{Miwa1999DX,
title={DX centers in GaAs/Si-$\delta$/AlAs heterostructure},
author={R. Miwa and T. M. Schmidt},
journal={Applied Physics Letters},
year={1999},volume={74},
pages={1999-2001},
doi={10.1063/1.123726}}

@article{Imamoglu_2021,
     author = {Atac Imamoglu and Ovidiu Cotlet and Richard Schmidt},
     title = {Exciton-polarons in two-dimensional semiconductors and the {Tavis-Cummings} model},
     journal = {Comptes Rendus. Physique},
     pages = {89--96},
     publisher = {Acad\'emie des sciences, Paris},
     volume = {22},
     number = {S4},
     year = {2021},
     doi = {10.5802/crphys.47},
}

@article{WHITTAKER19881,
title = {Theory of magneto-exciton binding energy in realistic quantum well structures},
journal = {Solid State Communications},
volume = {68},
number = {1},
pages = {1-5},
year = {1988},
issn = {0038-1098},
doi = {https://doi.org/10.1016/0038-1098(88)90233-5},
url = {https://www.sciencedirect.com/science/article/pii/0038109888902335},
author = {D.M. Whittaker and R.J. Elliott.},
}

@article{Bastard_1982,
  title = {Exciton binding energy in quantum wells},
  author = {Bastard, G. and Mendez, E. E. and Chang, L. L. and Esaki, L.},
  journal = {Phys. Rev. B},
  volume = {26},
  issue = {4},
  pages = {1974--1979},
  numpages = {0},
  year = {1982},
  month = {Aug},
  publisher = {American Physical Society},
  doi = {10.1103/PhysRevB.26.1974},
  url = {https://link.aps.org/doi/10.1103/PhysRevB.26.1974}
}

@article{Smith_1989,
  title = {Magnetoexciton spectrum of GaAs-AlAs quantum wells},
  author = {Smith, Doran D. and Dutta, M. and Liu, X. C. and Terzis, A. F. and Petrou, A. and Cole, M. W. and Newman, P. G.},
  journal = {Phys. Rev. B},
  volume = {40},
  issue = {2},
  pages = {1407(R)--1409(R)},
  numpages = {0},
  year = {1989},
  month = {Jul},
  publisher = {American Physical Society},
  doi = {10.1103/PhysRevB.40.1407},
  url = {https://link.aps.org/doi/10.1103/PhysRevB.40.1407}
}

@article{Potemski_1991,
  title = {Magnetoexcitons in narrow GaAs/${\mathrm{Ga}}_{1\mathrm{\ensuremath{-}}\mathit{x}}$${\mathrm{Al}}_{\mathit{x}}$As quantum wells},
  author = {Potemski, M. and Via, L. and Bauer, G. E. W. and Maan, J. C. and Ploog, K. and Weimann, G.},
  journal = {Phys. Rev. B},
  volume = {43},
  issue = {18},
  pages = {14707--14710},
  numpages = {0},
  year = {1991},
  month = {Jun},
  publisher = {American Physical Society},
  doi = {10.1103/PhysRevB.43.14707},
  url = {https://link.aps.org/doi/10.1103/PhysRevB.43.14707}
}

@article{Shields_1995,
  title = {Magneto-optical spectroscopy of positively charged excitons in GaAs quantum wells},
  author = {Shields, A.J. and Osborne, J.L. and Simmons, M.Y. and Pepper, M. and Ritchie, D.A.},
  journal = {Phys. Rev. B},
  volume = {52},
  issue = {8},
  pages = {R5523(R)--R5526(R)},
  numpages = {0},
  year = {1995},
  month = {Aug},
  publisher = {American Physical Society},
  doi = {10.1103/PhysRevB.52.R5523},
  url = {https://link.aps.org/doi/10.1103/PhysRevB.52.R5523}
}

@article{Finkelstein_1995,
  title = {Optical Spectroscopy of a Two-Dimensional Electron Gas near the Metal-Insulator Transition},
  author = {Finkelstein, Gleb and Shtrikman, Hadas and Bar-Joseph, Israel},
  journal = {Phys. Rev. Lett.},
  volume = {74},
  issue = {6},
  pages = {976--979},
  numpages = {0},
  year = {1995},
  month = {Feb},
  publisher = {American Physical Society},
  doi = {10.1103/PhysRevLett.74.976},
  url = {https://link.aps.org/doi/10.1103/PhysRevLett.74.976}
}

@article{Efimkin_2018,
  title = {Exciton-polarons in doped semiconductors in a strong magnetic field},
  author = {Efimkin, Dmitry K. and MacDonald, Allan H.},
  journal = {Phys. Rev. B},
  volume = {97},
  issue = {23},
  pages = {235432},
  numpages = {12},
  year = {2018},
  month = {Jun},
  publisher = {American Physical Society},
  doi = {10.1103/PhysRevB.97.235432},
  url = {https://link.aps.org/doi/10.1103/PhysRevB.97.235432}
}

@article{COMBESCOT2008215,
title = {The many-body physics of composite bosons},
journal = {Physics Reports},
volume = {463},
number = {5},
pages = {215-320},
year = {2008},
issn = {0370-1573},
doi = {https://doi.org/10.1016/j.physrep.2007.11.003},
url = {https://www.sciencedirect.com/science/article/pii/S0370157308000975},
author = {Monique Combescot and Odile Betbeder-Matibet and François Dubin},
}

@article{Villegas_Rosales_2022,
  title = {Composite fermion mass: Experimental measurements in ultrahigh quality two-dimensional electron systems},
  author = {Villegas Rosales, K. A. and Madathil, P. T. and Chung, Y. J. and Pfeiffer, L. N. and West, K. W. and Baldwin, K. W. and Shayegan, M.},
  journal = {Phys. Rev. B},
  volume = {106},
  issue = {4},
  pages = {L041301},
  numpages = {6},
  year = {2022},
  month = {Jul},
  publisher = {American Physical Society},
  doi = {10.1103/PhysRevB.106.L041301},
  url = {https://link.aps.org/doi/10.1103/PhysRevB.106.L041301}
}

@article{Kukushkin_2003,
author={Kukushkin, I. V.
and Smet, J. H.
and von Klitzing, K.
and Wegscheider, W.},
title={Cyclotron Resonance of Composite Fermions},
journal={Journal of Superconductivity},
year={2003},
month={Aug},
day={01},
volume={16},
number={4},
pages={777-781},
issn={1572-9605},
doi={10.1023/A:1025330310538},
url={https://doi.org/10.1023/A:1025330310538}
}

@article{Nicholas_1996,
doi = {10.1088/0268-1242/11/11S/003},
url = {https://doi.org/10.1088/0268-1242/11/11S/003},
year = {1996},
month = {nov},
publisher = {},
volume = {11},
number = {11S},
pages = {1477},
author = {R J Nicholas and D R Leadley and M S Daly and M van der Burgt and P Gee and J Singleton and D K Maude and J C Portal and J J Harris and C T Foxon},
title = {The dependence of the Composite Fermion effective mass on carrier density and Zeeman energy},
journal = {Semiconductor Science and Technology}
}

@article{Chughtai_2001,
title={Measurements of composite fermion masses from the spin polarization of two-dimensional electrons in the region 1},
author={R. Chughtai and V. Zhitomirsky and R. Nicholas and M. Henini},
journal={Phys. Rev. B},
year={2001},
volume={65},
pages={161305},
doi={10.1103/PhysRevB.65.161305}
}

@article{Kulah_2021,
title={The improved inverted AlGaAs/GaAs interface: its relevance for high-mobility quantum wells and hybrid systems},
author={E. K\"{u}lah and C. Reichl and J. Scharnetzky and L. Alt and W. Dietsche and W. Wegscheider},
journal={Semicond. Sci. Technol.},
year={2021},
volume={36},
pages={085013},
doi={10.1088/1361-6641/ac0d98}
}

\end{document}